\documentstyle[aps]{revtex}

\begin{document}
\title{Extending Sensitivity for Low-Mass Neutral Heavy Lepton Searches}
\author{Loretta M. Johnson and Douglas W. McKay}
\address{Department of Physics and Astronomy, University of Kansas, Lawrence, KS 66045}
\author{Tim Bolton}
\address{Department of Physics, Kansas State University, Manhattan, KS 66502}
\maketitle

\begin{abstract}
We point out the importance of two-body final states of weak isosinglet
neutral heavy leptons predicted in several models of new physics beyond the
standard model. We concentrate on muon-type neutral heavy leptons 
$L_\mu ^0$
with mass $M<2$ GeV which can be searched for with increased sensitivity at
a new round of neutrino experiments at CERN\ and Fermilab.  Providing
explicit decay rate formulae for the $e e \nu $, $e\mu \nu $, $\mu \mu 
\nu $, $\pi \mu $, $\rho \mu $, and $a_1\mu $ final states, we use 
general scaling features to estimate sensitivity of $L_\mu ^0$
searches in current and future experiments, emphasizing the importance of
the $\pi \mu $ decay mode.
\end{abstract}


\section{Introduction}

The LEP $N_\nu $ = 3 bound on the number of light, sequential-family
neutrinos~\cite{OPAL} tends to obscure the fact that relatively low-mass,
neutral leptons, $L^0$, weakly mixed with one or more of e, $\mu $, or $\tau 
$ neutrinos, are still allowed. Grand unified theory models provide
motivation, since some models may contain rather light neutral leptons~\cite
{wwlr}. Here we are specifically interested in $L^0$ with a mass $M$ of
order 1 GeV, and we assume for simplicity that a given massive neutrino
mixes with only one light neutrino flavor. Experimental bounds exist~\cite
{wr}, but it is timely to revisit them in light of the high intensity
neutrino experiments coming on line or under development and construction.

We especially emphasize the role of the $L^0\rightarrow
\pi ^{\pm }\ell ^{\mp }$ $\left( \ell ^{\mp }=e^{\mp },\mu ^{\mp }\right) $
decay mode in greatly increasing the sensitivity of searches for $L^0$ with $%
M<1$ GeV, where phase-space strongly favors this two-body mode. For $M>1$
GeV, the purely leptonic modes come into their own to provide clean signals
for $L^0$ decays.

%

\section{Heavy Neutral Lepton and Decays}

\subsection{Mixing}

In this study we are concerned with the search for heavy neutral leptons
which are primarily isosinglets under weak SU(2)$_L$. These can mix with the
light neutrinos, which are doublet members under SU(2)$_L$. Whether the
neutrinos are massless or massive but light,  the mixing effect can be
represented to a good approximation by replacing $\nu _{i_L},i=e,\mu ,\tau ,$
by 
\begin{equation}
\left( {\cal N}_i\right) _L\simeq \left( \nu _i\right) _L\left( 1-{\frac{%
\sum\limits_a|{\cal U}_{ia}|^2}2}\right) +\sum\limits_a{\cal U}_{ia}\left(
L_a^0\right) _L\;.  \label{Mixing}
\end{equation}
Here $\left( \nu _i\right) _L$ and $\left( L_a^0\right) _L$ are the
left--handed components of the neutrinos and heavy neutral leptons, while $%
\left( {\cal N}_i\right) _L$ are the combinations that appear in the charged
and neutral weak currents (the weak interaction ``eigenfields''). For
simplicity we consider the case where each $\nu _i$ mixes with only one $%
L_i^0$. If the neutrinos, $\nu _i$, are massive, the mixing can be
illustrated by a ``seesaw--like'' mass matrix\cite{seesaw}, 
\begin{equation}
{\cal M}=\left( 
\begin{array}{cc}
0 & \lambda  \\ 
\lambda  & M
\end{array}
\right) \;,
\end{equation}
where ${\cal M}$ has eigenvalues 
\begin{equation}
m_{\pm }=\frac M2\left( 1\pm \sqrt{1-{\frac{4\lambda }M}}\right) ,
\end{equation}
and corresponding eigenvectors 
\begin{equation}
v_{\pm }=\left( {\frac \lambda {\sqrt{m_{\pm }^2+\lambda ^2}}}\;,\;\pm {%
\frac{m_{\pm }}{\sqrt{m_{\pm }^2+\lambda ^2}}}\right) .
\end{equation}
When $\lambda /M<<1$, $L^0\equiv v_{+}\approx (\lambda /M,1-\lambda ^2/2M^2)$
is associated with $m_{+}\approx M$, while $\nu \equiv v_{-}\approx
(1-\lambda ^2/2M,-\lambda /M)$ belongs to $m_\nu \equiv m_{-}\approx \lambda
^2/M$. If, as in some specific models, $\lambda $ were related to charged
lepton or quark masses, the mixing parameter $U\simeq \lambda /M$ could be
determined by the heavy neutral lepton mass alone. 

In the work we present here, $U$ and $M$ are treated as independent
parameters. Since our treatment is perturbative, the mixing indicated in
Fig. \ref{MixFig} can be characterized by a mixing mass parameter $\lambda $%
, while the $L^0$ and $\nu $ lines represent propagators or a propagator and
an external wave function. For example, treating $L^0$ decay, one has
factors 
\begin{eqnarray}
\lambda \frac{\not{p}+m_\nu }{p^2-m_\nu ^2}U_{L^0}(p) &=&\frac \lambda {%
M-m_\nu }U_{L^0}(p) \\
&\simeq &\frac \lambda MU_{L^0}(p)  \nonumber
\end{eqnarray}
where $U_{L^0}(p)$ is the $L^0$ momentum space wave function and $m_\nu /M<<1
$. The parameters $\lambda $ and $M$ are to be determined, or bounded, by
experiment. 

\begin{figure}[tbph]
\caption{$L^0$ mixing. }
\label{MixFig}
\end{figure}

Figures \ref{FeynCC} and \ref{FeynNC} show the Feynman diagrams for the
charged and neutral current contributions to leptonic $L_a^0$ decay in
lowest order in $U_{ia}$'s and weak couplings.\footnote{%
For definiteness we diagram the decay as if the source beam were Dirac
NHL's. If the beam is sign-selected (like that of the NuTeV experiment, for
example), then this case applies directly. If the neutral lepton source is
from a beam dump (as in the DONUT experiment, for example), then there is a
democratic mixture of neutrino and antineutrino; and the conjugate $L^0$'s
and consequent conjugate lepton final states are all present. For
Majorana-type NHL, each charge state can occur in any decay regardless of
the nature of the beam preparation, and the diagram with the lepton switched
in Fig. \ref{MixFig} also occurs.} For a final state where $\ell _i\neq \ell
_j$ and both leptons are observed (which is required in the analysis
presented here), only the charged current processes, Fig. \ref{FeynCC},
contribute. Assuming the charges of the final leptons are determined, say $%
\ell _i^{-}$ and $\ell _j^{+}$, then there is a single diagram with factor $%
U_{ia}$. If the flavors, but not the charges, of the leptons are determined,
one has a second mode and another diagram with $\ell _j^{-}$ at the upper
vertex and $\ell _i^{+}$ at the lower one. The decay involves two terms, one
with a factor $|U_{ia}|^2$ and one with $|U_{ja}|^2$. For the e$\mu $ case, $%
\Gamma ^{e^{-}\mu ^{+}}=\Gamma ^{\mu ^{-}e^{+}}$, so $\Gamma ^{e\mu
}=(|U_e|^2+|U_\mu |^2)\Gamma ^{\mu ^{-}e^{+}}$.

\begin{figure}[tbph]
\caption{$L^0$ charged current decay. }
\label{FeynCC}
\end{figure}

\begin{figure}[tbph]
\caption{$L^0$ neutral current decay. }
\label{FeynNC}
\end{figure}

In the case that $\ell _i=\ell _j$, the neutral current diagrams are
involved, and each $\nu _i$ can contribute for a fixed $j$ and fixed $a$.
There is only {\it one} charged current contribution, proportional to $U_{aj}
$,  in this instance. The diagram where $i=j$ in Fig. \ref{FeynNC}, also
proportional to $U_{aj}$, interferes with that of Fig. \ref{FeynCC}, while
the other two terms with $i\neq j$ contribute incoherently to the partial
rate for $L_a^0\rightarrow \ell _j\bar{\ell}_j+$ missing neutrinos. We give
the expressions for the muon-type $L^0$ in Appendix \ref{Formulas}. The
numerical results and parameter bounds that we present will be for the
situation where only muon neutrinos mix with a given heavy neutral lepton.
The results can be generalized straightforwardly to the mixed cases. 

While on the topic of mixing, it is worth offering the reminder that heavy
neutral leptons with masses below $2$ GeV will live long enough\cite{Rosner}
to escape the LEP and SLC detectors before decaying and thus they are
disguised within the $N_\nu $ = 3 light-neutrino number measurements\cite
{OPAL}.  The right handed components of $L^0$ do not couple to $Z^0$, while
the left handed components couple through the mixing displayed in Eq. \ref
{Mixing} and in Fig. \ref{MixFig}, which ensures that the ``missing'' decay
partial width, which goes like $\sum\limits_{i=1}^3|\left( \bar{{\cal N}}%
_i\right) _L\left( {\cal N}_i\right) _L|^2$, gives the usual result as long
as $L^0$'s are light enough to be treated to the accuracy of present
measurements as massless particles in the final state. 

\subsection{Decays}

In this section, we present the partial decay widths of $L^0$ to channels of
relevance for our study. For definiteness we restrict our attention to a ``$%
\mu $--type'' $L_\mu ^0$ that mixes only with the weak muon neutrino with
mixing strength denoted by $U$.

\subsubsection{Leptonic Decays}

Referring to Fig. 2, we see that the relevant decays are 
\begin{equation}
L_\mu ^0\rightarrow \mu ^{-}e^{+}\nu _e
\end{equation}
and 
\begin{equation}
L_\mu ^0\rightarrow \mu ^{+}\mu ^{-}\nu _\mu \;.
\end{equation}
For convenience, we divide out a common factor 
\begin{equation}
K\equiv {\frac{G_F^2M^5}{192\pi ^3}}U^2\;,
\end{equation}
which happens to be the all-neutrino decay rate summed over flavor,
\begin{equation}
K=\sum_{i=e,\mu ,\tau }\Gamma \left( L_\mu ^0\rightarrow \nu _\mu \nu _i\bar{%
\nu}_i\right) ,
\end{equation}
and to the usual $\mu $--decay width formula when the electron mass is
neglected, up to the factor $U^2$. Differential and integrated partial width
formulas are summarized in Appendix \ref{Formulas}. Integrated partial
widths, $\Gamma (L_\mu ^0\rightarrow \mu ^{+}e^{-}\nu e)/K$ and $\Gamma
(L_\mu ^0\rightarrow \mu ^{+}\mu ^{-}\nu _\mu )/K$ are shown as a function
of $M$, the $L_\mu ^0$ mass, in Fig. \ref{PartialWidths}. Clearly the
three--body leptonic modes come into their own above $M=1$ GeV, but the
hadronic modes summarized next also play a major role in detection of $L_\mu
^0$ decay. 

\begin{figure}[tbph]
\caption{ The important leptonic and hadronic decay widths, $\Gamma/K$, 
as a function of M in GeV.  The solid curve is the $e \mu \nu $ mode, 
dotted is $\mu \mu \nu $, dashed is $\pi \mu $, dot-dashed is $\rho \mu 
$, and dot-dot-dashed is $a_1 \mu$.  Note that the $\pi \mu $ peak is 
more than 7 times higher than the $e \mu \nu $ peak. } 
\label{PartialWidths} 
\end{figure}

\subsubsection{Exclusive hadronic decays}

The exclusive decays 
\begin{equation}
L_\mu ^0\rightarrow \mu ^{-}H^{+},
\end{equation}
where $H=\pi ,\rho $ or $a_1$, are large modes that can be reliably
calculated within our framework. As we will explain in Sec. \ref{L0-expt},
only partial widths are needed for experimental searches since $U$ is
already constrained to be small. The signatures are clean, and the $L_\mu ^0$
detection sensitivity increases dramatically when these search modes are
included, as we discuss in the following section. The isospin-related
neutral modes, $L_\mu ^0\rightarrow \nu _\mu H^0$, are not as useful
experimentally, and are not explicitly included in the discussion. 

The $L_\mu ^0\rightarrow \mu ^{-}\pi ^{+}$ partial width is fixed by $\pi
^{+}\rightarrow \mu ^{+}\nu _\mu $ while $\Gamma \left( L_\mu ^0\rightarrow
\mu ^{-}\rho ^{+}\right) $ and $\Gamma \left( L_\mu ^0\rightarrow \mu
^{-}a_1^{+}\right) $ are fixed by $\tau ^{+}\rightarrow \rho ^{+}\nu _\tau $
and $\tau ^{+}\rightarrow a_1^{+}\nu _\tau $ decay. The ratios $\Gamma
(L_\mu ^0\rightarrow \mu ^{-}H^{+})/K$ are shown in Fig. \ref{PartialWidths}
as a function of $M$. The special prominence of the $L_\mu ^0\rightarrow \mu
^{-}\pi ^{+}$ mode for low mass $L_\mu ^0$ is evident. Regarding the
interpretation of the figure, note that two--body decays {\it appear} to
rise, peak and fall as a function of $M$ because of our normalization to the
3-neutrino decay width $K.$ Two body decays are proportional to $%
G_F^2M^3f_H^2$, where $f_H=f_\pi $ in the pion case, for example. The peak
in the curves for the $\pi ^{+},\rho ^{+}$ and $a_1^{+}$ final state partial
rates is then an artifact of the $M^{-2}$ from our normalization multiplied
by the rapidly rising two-body phase space.

As an aside, we note that two-body decays allow the potential discrimination
between electron-type and muon-type NHL through the observation of $L_\mu
^0\rightarrow \mu ^{-}H^{+}$ vs $L_e^0\rightarrow e^{-}H^{+}$.

Next we turn to a consideration of the issue of sensitivity to $L_\mu ^0$
detection in experimental searches, focussing on those aspects that follow
from basic scaling considerations and are independent of details of specific
experiments, which must be handled by thorough Monte Carlo analysis.

\section{Estimates of Sensitivity}

\subsection{Production of $L_\mu ^0$ from Meson Decay}

Light neutral heavy leptons could be produced in decays of any charged
stable meson; $K^{\pm }$, $D^{\pm }$, and $D_S^{\pm }$ provide the most
experimental sensitivity because they can be produced in enormous quantity
in the $pN$ collisions that generate neutrino beams and because CKM\
suppression of the two-body decay is modest($K^{\pm }$, $D^{\pm }$) or
absent ($D_S^{\pm }$). The two-body decay rate\cite{Rosner} can be related
directly to the $\mu \nu $ decay rate 
\begin{equation}
\Gamma (H\rightarrow \mu L_\mu ^0)=U^2\Gamma (H\rightarrow \mu \nu )\eta
_P(M/{M_H},{M_\mu }/M),  \label{Production}
\end{equation}
where $H=\left( K^{\pm },D^{\pm },D_S^{\pm }\right) ,$ $U^2$ is the mixing
factor, $M_H$ is the meson mass, $M$ is the $L_\mu ^0$ mass, $M_\mu $ is the
muon mass, and 
\begin{equation}
\eta _P(x,y)=\frac{\left[ \left( 1+y^2\right) -x^2\left( 1-y^2\right)
^2\right] \sqrt{\left[ 1-x^2\left( 1-y\right) ^2\right] \left[ 1-x^2\left(
1+y\right) ^2\right] }}{y^2\left( 1-x^2y^2\right) ^2}  \label{eta-def}
\end{equation}
is a kinematic factor. Note that for $M^2/M_\mu ^2\gg 1$, $\eta
_P(M/M_H,M_\mu /M)\rightarrow (M/M_\mu )^2(1-(M/M_H)^2)^2$, demonstrating 
that
two-body decays to $L_\mu ^0$ can be considerably enhanced by the lifting of
helicity suppression.

\subsection{\label{L0-expt}Experimental Detection of $L_\mu ^0$}

The simplest scheme for a detector is a low mass decay space immediately in
front of a neutrino detector\cite{Janet,CHARM NHL} instrumented with
tracking chambers to permit reconstruction of a possible $L_\mu ^0$ decay
vertex. By removing as much mass as possible, for example through the use of
helium-filled bags, backgrounds from conventional $\nu _\mu N$ 
interactions can be minimized. The neutrino detector downstream of the
low-mass region can be used to identify muons, pions, and electrons and
measure their energies, thus providing sensitivity to the largest decay
modes of $L_\mu ^0$ with $M<2$ GeV, $L_\mu ^0\rightarrow e\mu ,\mu \mu ,\pi
\mu $. While a dedicated instrumented low-mass detector is optimal, it is
not necessary, provided the density of material in front of the neutrino
detector is not too high.

If a decay space is constructed to have a length $\Delta $ along the
direction of the beam and a width much wider than the beam, the probability
of a $L_\mu ^0$ being observed in the detector can be expressed as 
\begin{equation}
P_D^{L_\mu ^0}(M,U^2)=\int_{Z-\Delta /2}^{Z+\Delta /2}\frac{dz}{\gamma \beta
c\tau _{L_\mu ^0}}e^{-z/\gamma \beta c\tau _{L_\mu ^0}}\frac{\Gamma _{L_\mu
^0}^{\det }}{\Gamma _{L_\mu ^0}^{tot}}\varepsilon _D^{L_\mu ^0},
\label{Exact Rate}
\end{equation}
where $\Gamma _{L_\mu ^0}^{\det },\Gamma _{L_\mu ^0}^{tot}$ are the detected
and total decay widths, $z$ is the distance of the decay position from the $%
L_\mu ^0$ production point, $Z$ is the distance from the production point to
the center of the decay channel, $\beta $ is the $L_\mu ^0$ speed in units
with $c=1$, $\gamma =1/\sqrt{1-\beta ^2}$, $\gamma \beta c\tau _{L_\mu ^0}$
is the $L_\mu ^0$ mean decay length, and $\varepsilon _D^{L_\mu ^0}$ is the
average $L_\mu ^0$ detection efficiency. Mixing angles of interest are
sufficiently small and detectors are configured such that $\Delta \ll Z\ll
\gamma \beta c\tau _{L_\mu ^0}$. In this case, Eq. \ref{Exact Rate}
simplifies to 
\begin{eqnarray}
P_D^{L_\mu ^0}(M,U^2) &=&\frac \Delta {\gamma \beta c\tau _{L_\mu ^0}}\frac{%
\Gamma _{L_\mu ^0}^{\det }}{\Gamma _{L_\mu ^0}^{tot}}\varepsilon _D^{L_\mu
^0}  \label{P-detect} \\
&=&\frac \Delta {\gamma \beta }\frac{\Gamma _{L_\mu ^0}^{\det }}{\hbar c}%
\varepsilon _D^{L_\mu ^0}.
\end{eqnarray}
Since $\tau _{L_\mu ^0}\Gamma _{L_\mu ^0}^{tot}=\hbar $, the observation
probability calculation requires only knowledge of the partial decay widths
of the channels being searched for, $\Gamma _{L_\mu ^0}^{\det }$. As these
partial widths can be reliably calculated, there should be little
theoretical uncertainty in estimates of search sensitivity.

\subsection{Sensitivity in $(U^2,M)$ Plane}

The number of produced $L_\mu ^0$ will be proportional to the total number
of protons on target, $N_{\text{POT}},$ and a factor $U^2M^2$ from Eq. \ref
{Production}, assuming $M_\mu ^2\ll M^2\ll M_H^2$. Using the simple $e\mu $
decay mode as an example, the detected partial width will be proportional to 
$U^2M^5$. The number of decays in the decay region is proportional to this
partial width, to the length of the decay space $\Delta $, and to an extra
factor of $M/E$ from time dilation, with $E$ the $L_\mu ^0$ energy.
Combining all effects together, the number of observed $L_\mu ^0$ has the
dependence 
\begin{equation}
N_{L_\mu ^0}^{obs}\propto \frac{N_{\text{POT }}\Delta }EU^4M^8\text{.}
\label{Sensitivity}
\end{equation}
If an experiment performs a search and observes a statistically significant $%
L_\mu ^0$ signal, then the experiment can determine both $M$, from the
two-body semihadronic decays, and $U^2$ from a more detailed development of
Eq. \ref{Sensitivity} (see Appendix \ref{NHL Sense}). On the other hand, if
no candidates are observed, then the experiment can exclude a region in the $%
(U^2,M)$ plane that from Eq. \ref{Sensitivity} will be bounded by a curve of
the form $U^2M^4=$ constant. The minimum mixing factor sensitivity for fixed 
$M$ will be proportional to $\sqrt{E/(N_{\text{POT }}\Delta )\text{ }}$ for
an experiment that suffers no background. Greater sensitivity follows from
increasing $N_{\text{POT }}$ and $\Delta $ or decreasing $E$, but the gain
is slow because of the square-root factor. Adding the hadronic decay
channels presented here is another way to improve the search limits.

\subsection{Estimates of Sensitivity for Current and Future Neutrino
Experiments}

This section summarizes predictions for two rather different experiments:
NuTeV\cite{NuTeV}, a high energy deep inelastic scattering experiment with
its neutrino detector located far from the production target, and DONUT\cite
{DONUT}, a high energy beam dump experiment with its detector 35 m from the
neutrino production target. We also comment on prospects in lower energy
experiments. A more detailed prescription for our estimates is given in
Appendix \ref{NHL Sense}.

\subsubsection{The NuTeV Experiment}

NuTeV\ has installed an instrumented decay channel to search for NHL\cite
{Janet} and is currently taking data. The 40 m long NHL decay channel is
located approximately 1200 m from the decay position for charmed mesons and
900-1200 m for charged kaons. NuTeV may receive an integrated intensity of
up to $6\times 10^{18}$ protons-on-target. Neutrino interactions occur at a
rate of approximately $20/10^{13}$ POT. The ratio of kaons to pions in the
NuTeV beam is about 0.4 resulting in $\nu _e/\nu _\mu $ interaction ratio of
about $2.3\%$ in the detector. Contributions from charmed meson decay
increase the $\nu _e$ rate by approximately $1\%$ of itself. The Lab E
neutrino detector used by NuTeV has a fiducial length of approximately 15 m
and a mean density of 4.2 g/cm$^3$.

Figure \ref{NHL K-rate} shows the estimated sensitivity plot in $U_K^2$ vs $%
M $ for NHL produced from $K^{\pm }$ decay in NuTeV. The main model
sensitivity comes from the assumption for the mean $K^{+}$ energy, which
enters both in the lifetime calculation of the $L_\mu ^0$ and the
interaction probability for the neutrinos used in the normalization sample.

\begin{figure}[tbph]
\caption{ 
Estimate of sensitivity as a function of mass $M$ (in GeV) to NHL produced 
from kaon decay in NuTeV. The vertical axis represents the minimum
mixing parameter $|U|^2$ the experiment would be sensitive to from
kaon decays alone assuming no candidate events were observed
(and no background events were expected) 
in an exposure to $6\times 10^{18}$ protons-on-target.
The dashed, solid, and dot-dashed curves are for kaon energies of 100, 150,
and 200 GeV, respectively, and assume the search is performed only using
the $L^0_\mu\rightarrow{e}\mu$ decay mode.  The dotted curve illustrates
the sensitivity that could be gained for 150 GeV kaon energy if the $\pi\mu$,
$\rho\mu$ and $a_1\mu$ modes were added to the $e\mu$ mode in the search.
}
\label{NHL K-rate}
\end{figure}
Figure \ref{NHL D-rate} shows a similar plot for NHL produced from $D^{\pm }_S$
decay. The main model sensitivity in this result comes from the assumption
for the mean $D^{+}_S$ energy. 
\begin{figure}[tbph]
\caption{ 
Estimate of sensitivity as a function of mass $M$ (in GeV) to NHL produced 
from $D_S$ decay in NuTeV. 
The vertical axis represents the minimum
mixing parameter $|U|^2$ the experiment would be sensitive to from
kaon decays alone assuming the no candidate events were observed
(and no background events were expected) 
in an exposure to $6\times 10^{18}$ protons-on-target.
The dashed, solid, and dot-dashed curves are for charmed hadron
 energies of 50, 100, and 200 GeV, respectively, 
and assume the search is performed only using
the $L^0_\mu\rightarrow{e}\mu$ decay mode.  The dotted curve illustrates
the sensitivity that could be gained for 100 GeV $D_S$ energy if the $\pi\mu$,
$\rho\mu$ and $a_1\mu$ modes were added to the $e\mu$ mode in the search.
}
\label{NHL D-rate}
\end{figure}

Figure \ref{NHL estimates} shows the estimated sensitivity for all modes
combined using either the $e\mu $ decay channel by itself or the $e\mu $ and 
$\pi \mu $ channels combined. Note that the $U^4$ dependence of the
detection rate for NHL implies that uncertainties in acceptances, branching
fractions, etc., which only enter as ratios, affect results only as square
roots in determining $U^2$. The dependence of the result on statistics,
assuming no background, is similarly proportional to $1/\sqrt{N_{\text{POT}}}
$.

\begin{figure}[tbph]
\caption{ 
Estimate of combined sensitivity to NHL from 
$K$, $D$, and $D_S$ decays in
NuTeV as a function of NHL mass $M$ (in GeV).
The vertical axis represents the minimum
mixing parameter $|U|^2$ the experiment would be sensitive to from
all meson decays combined assuming no candidate events were observed
(and no background events were expected).  
The curves assume $6\times 10^{18}$ POT, a mean kaon energy
of 150 GeV, and a mean charmed meson energy  of 100 GeV.
The solid curve shows the expected result 
assuming the search were performed only using
the $L^0_\mu\rightarrow{e}\mu$ decay mode.  
The dotted curve illustrates
the sensitivity that could be gained  by adding the $\pi\mu$,
$\rho\mu$, and $a_1\mu$ modes to the search.
}
\label{NHL estimates}
\end{figure}

\subsubsection{The DONUT\ Experiment}

DONUT\ is a hybrid emulsion spectrometer detector sited approximately 35 m
from a beam dump target in Fermilab's 800 GeV\ proton beam. Active and
passive shielding eliminate essentially all neutrinos produced by pion and
kaon decay, leaving a mixed beam of $\nu _e$, $\nu _\mu $, and $\nu _\tau $
from charmed hadron decay. The experiment's primary goal is to detect
charged current interactions of $\nu _\tau $ with nucleons in the emulsion
target. DONUT may receive an exposure of up to $2\times 10^{18}$
protons-on-target.

DONUT's proximity to the production target greatly enhances the flux of
neutrinos, and possible NHL, produced from $D^{\pm }$ and $D_S^{\pm }$
decay. Only a few thousand charged current interactions in the emulsion
detector are expected, but these will essentially all originate from charm
decay. Figure \ref{DONUT NHL} shows an estimate to sensitivity\cite{Noel}
for NHL\ in DONUT, assuming the experiment could instrument a 5 m decay
space in front of their emulsion target, and that the experiment receives an
exposure of $2\times 10^{18}$ POT. The estimate is comparable to that for
NuTeV, and possibly better at high $L_\mu ^0$ mass.

\begin{figure}[tbph]
\caption{ 
Estimate of combined sensitivity to NHL from 
$D$ and $D_S$ decays in
DONUT as a function of NHL mass $M$ (in GeV).
The vertical axis represents the minimum
mixing parameter $|U|^2$ the experiment would be sensitive to from
all meson decays combined assuming the no candidate events were observed
(and no background events were expected).  
The curves assume $2\times 10^{18}$ POT and 
a mean charmed meson energy  of 100 GeV.
The solid curve shows the expected result 
assuming the search were performed only using
the $L^0_\mu\rightarrow{e}\mu$ decay mode.  
The dotted curve illustrates
the sensitivity that could be gained  by adding the $\pi\mu$,
$\rho\mu$, and $a_1\mu$ modes to the search.
}
\label{DONUT NHL}
\end{figure}

\subsubsection{Other Experiments}

We have used the NuTeV and DONUT\ experiments as specific examples in
calculating sensitivity to NHL; however, our formulas can easily be applied
in other cases. CHORUS\cite{CHORUS}, also a hybrid emulsion-spectrometer,
and NOMAD\cite{NOMAD}, a low mass high resolution spectrometer, are running
in a low energy horn beam at CERN to search for $\nu _\tau $ produced from $%
\nu _\mu \rightarrow \nu _\tau $ oscillations. Two additional $\nu _\mu
\rightarrow \nu _\tau $ oscillation experiments, COSMOS\cite{COSMOS} and
MINOS\cite{MINOS}\ at Fermilab, will begin taking data around 2001. All of
these experiments can search for NHL produced from (primarily) kaon decay
because of their lower energy beams. We note that searches in this mass
regime benefit appreciably from an ability to detect the $\pi \mu $ decay
mode.

\section{Summary}

As just discussed, a number of Fermilab and CERN neutrino oscillation
experiments are currently running or will soon come on-line, and we have
shown that a simple and direct method to expand the search for light (M $<$
2 GeV) neutral heavy leptons (NHL) is suitable for all of these experiments.
The present lower limits on lifetimes for NHL's in this mass range means
that all experiments satisfy the criterion that $\Delta <<z<<\gamma \beta
c\tau _{L^0}$ where $\Delta $ is the fiducial length, z is the
source-to-detector distance and $\gamma \beta c\tau $ is the decay length.
Therefore the simple criterion (\ref{P-detect}) applies. Only theoretical
values of the partial widths into search modes are relevant, and these can
be reliably calculated in terms of the mass and mixing parameters.

In addition to common features that lend themselves to a clean analysis, we
have shown how important the two-body decay modes are, especially the $\pi
\mu $ mode, in achieving improvements in sensitivity. We found significant
gains over most of the mass range $M<2$ GeV, which means discovery reach
extends to smaller mixing values than currently reported in the literature.

Individual experiments will use sophisticated Monte Carlo simulations  to
properly account for details such as  efficiencies and fluxes;  but the
simple and general formalism we have presented applies to all experiments.
We believe our results provide a broad, useful framework for expanding
searches for  neutral heavy leptons.

\section*{Acknowledgments}
TB would like to thank Janet Conrad for numerous discussions about the 
NuTeV decay channel project.  LMJ acknowledges support from a University 
of Kansas Graduate Summer Fellowship.  DWM thanks S. Ranjbar-Daemi and 
Faheem Hussain for the hospitality of the high energy group at ICTP, 
Trieste.  This work was supported in part by DOE grants \# 
DE-FG02-85ER40214 and DE-FG02-94ER40814.

\appendix 

\section{Explicit Decay Rate Formulae}

\label{Formulas}

In this appendix we collect formulas for partial decay widths to which we
refer in the text. For completeness we also include the differential forms
before the final integrations, including effects of $L^0$ polarization in 
the $e\mu $ decay channel. All formulas refer to the $L^0$ rest frame.

\subsection{Leptonic Rates}

Referring to Fig. {\ref{FeynCC}} and identifying $\mu ^{-}=\ell _i$ and $%
e^{+}=\bar{\ell}_j$, with $x_\mu \equiv 2E_\mu /M$, $x_e\equiv 2E_e/M$, and $%
x_m\equiv M_\mu /M$, we have 
\begin{equation}
\frac{d^2\Gamma ^{(\mu ^{-}e^{+})}}{dx_edx_\mu }=\frac{G_F^2M^5}{16\pi ^3}%
|U_\mu |^2[x_e(1-x_e-x_m^2)]\;,  \label{mue2}
\end{equation}
where $M$ is the $L^0$ mass, $M_\mu $ is the muon mass and $E_e$ and $E_\mu $
are the positron and muon energies. We have neglected the electron mass in
writing Eq. \ref{mue2}. Saving the $x_\mu $ integration until last, and
setting $m_e=0$ in the phase--space treatment, the single differential decay
form is given by the expression 
\begin{equation}
\frac{d\Gamma ^{(\mu ^{-}e^{+})}}{dx_\mu }=\frac{G_F^2M^5}{16\pi ^3}|U_\mu
|^2\left[ -{\frac 23}x_m^2+{\frac 12}(1+x_m^2)x_\mu -{\frac 13}x_\mu
^2\right] \sqrt{x_\mu ^2-4x_m^2}\;.  \label{mue1}
\end{equation}
Integrating over $x_\mu $ from  $2x_m\leq x_\mu \leq 1+x_m^2$, we get the
result familiar from muon decay, 
\begin{equation}
\Gamma ^{(\mu ^{-}e^{+})}=\frac{G_F^2M^5}{192\pi ^3}|U_\mu
|^2(1-8x_m^2+8x_m^6-x_m^8-12x_m^4\ln x_m^2)\;.  \label{mue}
\end{equation}

For completeness we include the corresponding expressions for the $%
L_e^0\rightarrow e^{-}\mu ^{+}\nu _\mu $ final state. 
\begin{equation}
\frac{d^2\Gamma ^{(e^{-}\mu ^{+})}}{dx_edx_\mu }=\frac{G_F^2M^5}{16\pi ^3}%
|U_e|^2[x_\mu (1-x_\mu +x_m^2)]\;,  \label{emu2}
\end{equation}
where again the electron mass is neglected. Integrating over the electron
energy first, we obtain 
\begin{equation}
{\frac{d\Gamma ^{(e^{-}\mu ^{+})}}{dx_\mu }}=\frac{G_F^2M^5}{16\pi ^3}%
|U_e|^2x_\mu (1-x_\mu +x_m^2)\;\sqrt{x_\mu ^2-4x_m^2}\;.  \label{emu1}
\end{equation}
Comparing Eq.'s \ref{mue1} and {\ref{emu2}}, we see the interesting feature
that, in principle, the difference in the two distributions could
distinguish $L_\mu ^0\rightarrow \mu ^{-}e^{+}\nu _e$ from $L_e^0\rightarrow
e^{-}\mu ^{+}\nu _\mu $ even if the charge states were not determined
directly. This is true even if $x_m<<1$, though distinguishing between
distributions $\Gamma ^{(\mu ^{-}e^{+})}\sim x_\mu ^2-{\frac 23}x_\mu ^3$
and $\Gamma ^{(e^{-}\mu ^{+})}\sim x_\mu ^2-x_\mu ^3$ may be difficult.
Integrating over the muon energy, one finds Eq. \ref{mue}.

If the $L_\mu ^0$ is produced through meson decay, it will likely be highly
polarized. Correlations between the polarization vector and the $e$ and $\mu 
$ momentum vectors could produce substantial effects in detection efficiency
for the $L^0\rightarrow \mu ^{-}e^{+}\nu _e$ final state. To permit study of
these effects, we give the following triply differential decay formulas: 
\begin{equation}
\frac{d^3\Gamma ^{(\mu ^{-}e^{+})}}{dx_\mu dx_edcos\theta _e}=\frac{%
G_F^2 M^5}{32\pi ^3}|U_\mu |^2 x_e(1-x_m^2-x_e)(1-\cos \theta _e)
\end{equation}
where $\theta _e$ is the angle between the final $e^{+}$ direction and the
polarization direction of the decaying $L^0$,and $m_e$ is neglected as
before; or, alternatively, 
\begin{equation}
\frac{d^3\Gamma ^{(\mu ^{-}e^{+})}}{dx_\mu dx_edcos\theta _\mu }=\frac{%
G_F^2 M^5}{32\pi ^3}|U_\mu |^2x_e(1-x_m^2-x_e)(1-\cos \tilde{\theta}_e\cos 
\theta _\mu ),
\end{equation}
where

\begin{equation}
\cos \tilde{\theta}_e=\frac{(2-x_\mu -x_e)^2-x_\mu ^2-x_e^2+4x_m^2}{2x_e%
\sqrt{x_\mu ^2-4x_m^2}}
\end{equation}
and $\theta _\mu $ is the angle between the final $\mu ^{-}$ direction and
the polarization direction of the decaying $L^0$.

Turning to the case where there are two muons in the final state and the
charged and neutral current contributions interfere, we write the double
differential decay formula as 
\begin{eqnarray}
\frac{d^2\Gamma ^{(\mu ^{-}\mu ^{+})}}{dx_{+}dx_{-}} &=&{}\frac{G_F^2M^5}{%
64\pi ^3}|U_\mu |^2[x_{+}(1-x_{+})(a+b)^2+x_{-}(1-x_{-})(a-b)^2
\label{mumu2} \\
&&+2x_m^2(2-x_{-}-x_{+})(a^2-b^2)]\;,  \nonumber
\end{eqnarray}
where $x_{\pm }=2E_{\mu ^{\pm }}/M$, $a=(3/2-2\sin ^2\theta _W)$ and $b={1/2}
$, and $\sin ^2\theta _W\cong 0.224$. Integrating over $x_{+}$, one obtains 
\begin{eqnarray}
\frac{d\Gamma ^{(\mu ^{-}\mu ^{+})}}{dx_{-}} &=&\frac{G_F^2M^5}{16\pi ^3}%
|U_\mu |^2\{[x_{-}(1-x_{-})(a-b)^2  \nonumber \\
&&+2x_m^2(2-x_{-})(a^2-b^2)]\;(x_{-}^u-x_{-}^\ell )  \nonumber \\
&&+{\frac 12}[(a+b)^2-2x_m^2(a^2-b^2)]\;[(x_{-}^u)^2-(x_{-}^\ell )^2] 
\nonumber \\
&&-{\frac 13}(a+b)^2[(x_{-}^u)^3-(x_{-}^\ell )^3]\;.  \label{mumu1}
\end{eqnarray}
The expressions for $x_{-}^u$ and $x_{-}^\ell $ in Eq. \ref{mumu1} read: 
\begin{equation}
x_{-}^{u,\ell }=\frac{[-({\frac{x_{-}}2}-1)(1-x_{-}+2x_m^2)\pm {\frac 12}%
\sqrt{x_{-}^2-4x_m^2}(1-x_{-})]}{(1+x_m^2-x_{-})}\;.  \label{limits}
\end{equation}
One obtains the corresponding expression for $\frac{d\Gamma ^{(\mu ^{-}\mu
^{+})}}{dx_{+}}$ by the replacement $x_{-}\rightarrow x_{+}$. Integrating
the expression \ref{mumu1} over $x_{-}$, one obtains the partial width
formula for $L^0\rightarrow \mu ^{-}\mu ^{+}\nu _\mu $; namely 
\begin{eqnarray}
\Gamma ^{(\mu ^{-}\mu ^{+})} &=&\frac{G_F^2M^5}{192\pi 
^3}|U_\mu |^2 \left\{
C_1\left[ (1-14x_m^2-2x_m^4-12x_m^6)\sqrt{1-4x_m^2}\right. \right.  
\nonumber \\
&&\left. +12x_m^4(x_m^4-1)L\right] +4C_2\left[ x_m^2(2+10x_m^2-12x_m^4)\sqrt{%
1-4x_m^2}\right.   \nonumber \\
&&\left. \left. +6x_m^4(1-2x_m^2+2x_m^4)L\right] \right\} \;,  \label{mumu}
\end{eqnarray}
where 
\begin{equation}
L=\log \left[ \frac{1-3x_m^2-(1-x_m^2)\sqrt{1-4x_m^2}}{x_m^2(1+\sqrt{1-4x_m^2%
})}\right] \;.  \label{log}
\end{equation}
The coefficients $C_1$ and $C_2$ in Eq. \ref{mumu} are 
\begin{eqnarray}
C_1 &=&\frac 14\left[ 1+4\sin ^2\theta _W+8\sin ^4\theta _W\right] , \\
C_2 &=&\frac 12\sin ^2\theta _W[1+2\sin ^2\theta _W]\;.
\end{eqnarray}

For completeness we give the corresponding $L_\mu^0\rightarrow e^+ e^- 
\nu_\mu$ partial width 

\begin{equation}
\Gamma ^{(e ^{-} e ^{+})} = \frac{G_F^2M^5}{192\pi 
^3}|U_\mu |^2 \frac{1}{4} \left[ 1-4\sin ^2\theta_W+8\sin 
^4\theta_W\right] .
\end{equation}

\subsection{Two--body Semihadronic Decay Formulas}

The three decay modes of interest in this work are $L^0\rightarrow \pi
^{+}\mu ^{-},\rho ^{+}\mu ^{-}$ and $a_1^{+}\mu ^{-}$, which involve the
decay constants $f_\pi ,$ $g_\rho $, and $g_a$. We determine $g_\rho $ and $%
g_a$ from the partial widths for $\tau \rightarrow \rho \nu _\tau $ and $%
\tau \rightarrow a_1\nu _\tau $ using the lowest order diagrams. As
justification, we note that the $\tau \rightarrow \pi \nu _\tau $ partial
width calculated with tusing the measured value of $f_\pi $ agrees with the
experiment to within a few percent. The decay formulas in the rest frame of $%
L^0$ read, in the narrow $\rho ,a_1$ width approximation 
\begin{equation}
{\frac{d\Gamma ^{(\pi \mu )}}{d\Omega }}=f_\pi ^2{\frac{cos^2\theta _c}{%
64\pi ^2}}G_F^2|U_\mu |^2M^3\sqrt{{\cal S}(M,M_\pi ,M_\mu )}\left[ \left( 1-{%
\frac{M_\mu ^2}{M^2}}\right) ^2-{\frac{M_\pi ^2}{M^2}}\left( 1+{\frac{M_\mu
^2}{M^2}}\right) \right] ,  \label{pil}
\end{equation}

\begin{equation}
{\frac{d\Gamma ^{(\rho \mu )}}{d\Omega }}={\frac{g_\rho ^2}{M_\rho ^2}}{%
\frac{cos^2\theta _c}{32\pi ^2}}G_F^2|U_\mu |^2M^3\sqrt{{\cal S}(M,M_\rho
,M_\mu )}\left[ \left( 1+{\frac{M_\mu ^2}{M^2}}\right) {\frac{M_\rho ^2}{M^2}%
}-2{\frac{M_\rho ^4}{M^4}}+\left( 1-{\frac{M_\mu ^2}{M^2}}\right) ^2\right] ,
\label{rhol}
\end{equation}
where 
\begin{equation}
{\cal S}(M,M_H,M_\mu )=\left[ 1-\left( {\frac{M_H}M}-{\frac{M_\mu }M}\right)
^2\right] \left[ 1-\left( {\frac{M_H}M}+{\frac{M_\mu }M}\right) ^2\right] 
\end{equation}
and $d\Gamma ^{(a_1\ell )}/d\Omega $ is obtained from Eq. \ref{rhol} with
the replacements $g_\rho \rightarrow g_a$ and $M_\rho \rightarrow M_a$. The
integrated partial widths are obviously obtained by multiplying Eq.'s \ref
{pil} and \ref{rhol} by $4\pi $.

The parameters $g_\rho $ and $g_a$ are determined from the $\tau $ partial
widths\cite{1996 PDG} to be 
\begin{eqnarray}
g_\rho ^2 &=&\frac{\Gamma (\tau \rightarrow \rho \nu _\tau )}{{\frac{G_F^2}{%
8\pi }}cos^2\theta _c{\frac{M_\tau ^3}{M_\rho ^2}}\left( 1-{\frac{M_\rho ^2}{%
M_\tau ^2}}\right) \left( 1+{\frac{2M_\rho ^2}{M_\tau ^2}}\right) } 
\nonumber \\
&=&(0.102\text{ GeV}^2)^2,  \label{grho}
\end{eqnarray}
and similarly 
\begin{equation}
g_a^2=\left( 0.128\text{ GeV}^2\right) ^2\;.  \label{ga}
\end{equation}
The branching fraction for $\tau ^{-}\rightarrow \pi ^{-}\pi ^{+}\pi
^{-}+\geq 0$ neutrals $+\nu _\tau $ was used for the $a_1$ fraction. A
recent chiral dynamics analysis of mesons\cite{chiral} yields 
\begin{equation}
g_\rho =0.104\text{ GeV}^2
\end{equation}
and 
\begin{equation}
g_a=0.136\text{ GeV}^2\;,
\end{equation}
in reasonable agreement with the values in Eq.'s \ref{grho} and \ref{ga}.

\section{Detailed Estimates of Sensitivity}

\label{NHL Sense}

A useful way to estimate sensitivity to NHL couplings is to normalize to
neutrino interactions in the detector produced by the same parent meson as
that which produces the NHL. This technique relies less on absolute
calculations from a beam monte carlo and allows one to identify important
model dependencies.

This section will develop estimates of sensitivity to NHL produced from $%
K^{\pm },D^{\pm }$, and $D_S^{\pm }$ two-body meson decays. At the end we
comment on NHL search possibilities in all current and future accelerator
based neutrino experiments that we are aware of. 

\subsection{$\mu $-type NHL from Kaons}

We define the observable 
\begin{eqnarray}
R_K(M,U^2) &=&\frac{\mbox{Detected }L_\mu ^0\mbox{ in Channel from }K^{+}%
\mbox{ Decays }}{\mbox{Detected }\nu _\mu \mbox{ in Detector from }K^{+}%
\mbox{ Decays }} \\
&=&R_P(M,U^2)R_B(M,U^2)R_D(M,U^2),  \nonumber
\end{eqnarray}
with $R_P(M,U^2)$, $R_B(M,U^2)$, and $R_D(M,U^2)$  defined as the $L_\mu ^0$
production ratio, beam transport ratio, and detection ratio, respectively.
The number of detected $L_\mu ^0$ will be a strong function of $L_\mu ^0$
mass $M$ and mixing factor $U^2$. The number of ordinary $\nu _\mu $
detected from $K^{+}$ decays can be inferred from the energy spectrum of
neutrino interactions in the neutrino detector. By formulating the search as
a measurement of the ratio $R_K(M,U^2)$, one lessens sensitivity to absolute
normalization of the neutrino beam and detector acceptance. To establish the
limits of experimental sensitivity, we will consider the null case, where no 
$L_\mu ^0$ candidates are observed for a given exposure in a neutrino beam.
In this case, the upper 90\% confidence level limit sensitivity for $%
R_K(M,U^2)$ follows, assuming no observed events and no background, from
Poisson statistics, 
\begin{equation}
R_K(M,U^2)\leq \frac{2.3}{r_{K\pi }n_{\nu _\mu }N_{\mbox{POT }}},
\label{K-expt}
\end{equation}
with $r_{K\pi }$ the fraction of $\nu _\mu $ events from kaons, $n_{\nu _\mu
}$ the number of $\nu _\mu $ interactions in the detector per incident
proton, and $N_{\mbox{POT }}$the total number of protons on the production
target. This section will derive the functional dependence of $R_K(M,U^2)$
on $M$ and $U^2$ that will allow limits to be placed on the $(M,U^2)$ plane
from a null result search.

\subsubsection{Production Factor $R_P(M,U^2)$}

$R_P(M,U^2)$ is the ratio of produced NHL to produced $\nu _\mu $ from kaons%
\cite{Rosner}: 
\begin{equation}
R_P(M,U^2)=\frac{U^2M^2}{M_\mu ^2}\frac{\left( 1-\frac{M^2}{M_K^2}\right) ^2%
}{\left( 1-\frac{M_\mu ^2}{M_K^2}\right) ^2}\eta _P(\frac M{M_K},\frac{M_\mu 
}M),
\end{equation}
where, as before, $U$ is the $\nu _\mu -L_\mu ^0$ mixing strength, $\eta _P$
is defined in Eq. \ref{eta-def},  and $M$, $M_\mu $, and $M_K$ are the NHL,
muon, and charged kaon masses, respectively. If the muon mass is neglected,
the phase space function $\eta _P(x,y)$ is unity.

\subsubsection{Beam Factor $R_B(M,U^2)$}

$R_B(M,U^2)$ is the relative acceptance for NHL vs neutrinos due to the
beamline. An experiment will generally run a detailed Monte Carlo simulation
to obtain this factor, but at high energies, $E_{L_\mu ^0}\gg M$, one would
expect only modest acceptance differences. Accordingly, we assume for
estimation purposes that 
\begin{equation}
R_B(M,U^2)=\frac{\varepsilon _B^{L_\mu ^0}}{\varepsilon _B^{\nu _K}}\approx
1.0.
\end{equation}

\subsubsection{Detection Factor $R_D(M,U^2)$}

$R_D(M,U^2)$ is the relative detection probability for NHL vs $\nu _\mu $.
This can be expressed as a ratio of detection probabilities, 
\begin{equation}
R_D(M,U^2)=\frac{P_D^{L_\mu ^0}(M,U^2)}{P_D^{\nu _K}}.
\end{equation}
For long NHL lifetime, the NHL detection probability $P_D^{L_\mu ^0}(M,U^2)$
can be written   
\begin{equation}
P_D^{L_\mu ^0}(M,U^2)=\frac \Delta {\gamma \beta c\tau _{L_\mu ^0}}\frac{%
\Gamma _{L_\mu ^0}^{\det }}{\Gamma _{L_\mu ^0}^{tot}}\varepsilon _D^{L_\mu
^0},
\end{equation}
where $\Gamma _{L_\mu ^0}^{\det },\Gamma _{L_\mu ^0}^{tot}$ are the detected
and total decay widths, $\Delta $ is the length of the decay channel, $%
\gamma \beta c\tau _{L_\mu ^0}$ is the $L_\mu ^0$ decay length, and $%
\varepsilon _D^{L_\mu ^0}$ is the mean $L_\mu ^0$ detection efficiency, 
assumed to be 1.0.
Choosing to observe only the $e\mu $ mode for simplicity, $P_D^{L_\mu
^0}(M,U^2)$ can be re-expressed as 
\begin{equation}
P_D^{L_\mu ^0}(M,U^2)=\frac{U^2M^6\Delta }{\left\langle E_{L_\mu
^0}\right\rangle M_\mu ^5c\tau _\mu }\varepsilon _D^{L_\mu ^0}\eta _D(M_\mu
/M),
\end{equation}
with $\eta _D\left( \frac{M_\mu }M\right) $ a threshold factor in the $L_\mu
^0$ decay that can be read from Eq. \ref{mue} for the $e\mu $ decay.

The neutrino interaction probability is simply 
\begin{equation}
P_D^{\nu _K}(M,U^2)=N_A\left\langle \rho t\right\rangle \sigma _\nu ^{\prime
}\left\langle E_\nu \right\rangle \varepsilon _D^{\nu _K},
\end{equation}
where $\sigma _\nu ^{\prime }\left\langle E_\nu \right\rangle =0.68\times
10^{-38}\left\langle E_\nu \right\rangle $ cm$^2$ is the neutrino-nucleon
cross section, $N_A\left\langle \rho t\right\rangle $ is the target
thickness in units of nucleons/cm, and $\varepsilon _D^{\nu _K}$ is the
target detection efficiency, which is usually close to $1.0$.

\subsubsection{Sensitivity Formula}

The final prediction for $R_K(M,U^2)$ is of the form 
\begin{equation}
R_K(M,U^2)=CU^4M^8\left( 1-\frac{M^2}{M_K^2}\right) ^2,  \label{K-theory}
\end{equation}
with $C$ containing all experiment-specific information, 
\[
C=a\frac{\eta _P(\frac M{M_K},\frac{M_\mu }M)\eta _D\left( \frac{M_\mu }M%
\right) \varepsilon _D^{L_\mu ^0}\Delta }{\left\langle E_{L_\mu
^0}\right\rangle \left\langle E_\nu \right\rangle \varepsilon _D^{\nu
_K}\left\langle \rho t\right\rangle }, 
\]
and $a$ a constant, 
\begin{eqnarray}
a &=&\left[ M_\mu ^7\left( 1-\frac{M_\mu ^2}{M_K^2}\right) ^2c\tau _\mu
N_A\sigma _\nu ^{\prime }\right] ^{-1} \\
&=&2.7\times 10^{16}\mbox{ cm}^{-3}\mbox{g}^{-1}\mbox{GeV}^{-6}.  \nonumber
\end{eqnarray}
For fixed $M$, it follows from combining equation \ref{K-theory} with
equation \ref{K-expt} that 
\begin{equation}
U_K^2\leq M^{-4}\sqrt{\frac 1C\left( 1-\frac{M^2}{M_K^2}\right) ^{-1}\frac{%
2.3}{r_{K\pi }n_{\nu _\mu }N_{\mbox{POT }}}}.  \label{K-sense}
\end{equation}
This formula demonstrates the qualitative feature of any NHL limit based on
a search for decay vertices of NHL produced from meson decay. Sensitivity to
the mixing $U_K^2$ is proportional to $M^{-4}$ up to masses near the
kinematic limit. The $M^{-4}$ behavior originates from the $U^4M^8$
dependence of $R_K(M,U^2)$, with $U^2M^2$ arising from NHL production, $%
U^2M^5$ from NHL\ decay, and an extra factor of $M$ from time dilation.
Because of the $U^4$ dependence of $R_K(M,U^2)$, limits on $U^2$ only
improve as the square root of the total integrated protons on target. The
square root also insures that other experimental effects, which enter as
ratios in $C$, only weakly affect the sensitivity estimate. 

\subsection{$\mu $-type NHL from D$^{+}$ Mesons}

The calculation proceeds in a manner similar to the kaon case, except this
time a more convenient normalization is relative to electron-neutrinos from
charmed mesons: 
\begin{eqnarray}
R_C(M,U^2) &=&\frac{\mbox{Detected }L_\mu ^0\mbox{ in Channel from }D^{+}%
\mbox{ Decays }}{\mbox{Detected }\nu _e\mbox{ in Detector from }D^{+}%
\mbox{
Decays }} \\
&=&R_P^{\prime }(M,U^2)R_B^{\prime }(M,U^2)R_D^{\prime }(M,U^2).  \nonumber
\end{eqnarray}
This choice is motivated by the fact that charm decays produce a negligible
fraction of the total $\nu _\mu $, but on the order of $1\%$ of the $\nu _e$
in a high energy beam. The 90\% C.L. Poisson sensitivity to $R_C(M,U^2)$ can
be written as 
\begin{equation}
R_C(M,U^2)\leq \frac{2.3}{f_{\nu _eD}f_{\nu _e}n_{\nu _\mu }N_{\mbox{POT }}},
\nonumber
\end{equation}
where $f_{\nu _e}$ is the fraction of neutrino events which are $\nu _e$ and 
$f_{\nu _eD}$ is the fraction of interacting electron neutrinos from $D^{+}$
decay. Note that $f_{\nu _eD}\rightarrow 1$ in an ideal beam-dump
experiment, whereas $f_{\nu _eD}\simeq 0.01$ in a conventional high energy
experiment.

For the production rate, one can assume that all conventional $\nu _e$ are
produced from three-body semileptonic decays whereas the NHL production is
dominantly two-body, 
\begin{equation}
R_P^{\prime }(M,U^2)=\frac{U^2M^2}{M_\mu ^2}{\left( 1-\frac{M^2}{M_D^2}%
\right) ^2}\frac{B(D^{+}\rightarrow \mu ^{+}\nu _\mu )}{B(D^{+}\rightarrow
\mu ^{+}X)}.
\end{equation}
In this formula the muon mass is neglected compared to the $D^{+}$ mass $M_D$%
, and the ratio $B(D^{+}\rightarrow \mu ^{+}\nu _\mu )/B(D^{+}\rightarrow
\mu ^{+}X)$ corrects for the use of three body decays of the $D^{+}$ as the
dominant source of $\nu _e$. The $D^{+}$ muonic rate can be calculated
assuming a $D$ meson decay constant $f_D$\cite{Pat and Jeff} via $%
B(D^{+}\rightarrow \mu ^{+}\nu _\mu )=(3.2\times 10^{-4})\left( \frac{f_D}{%
200\mbox{ MeV}}\right) ^2$, and the semi-muonic rate $B(D^{+}\rightarrow \mu
^{+}X)$ is measured as $B(D^{+}\rightarrow \mu ^{+}X)=0.172$. 

For the beam transport factor, we assume again a value of 
\begin{equation}
R_B^{\prime }(M,U^2)=1.0.
\end{equation}
This number may be less than $1.0$ owing to the higher $p_T$ given to the $%
L_\mu ^0$ relative to the three-body decay neutrino.

The detection factor is modified by the softer $D^{+}$ spectrum, relative to
kaons, and the somewhat higher $L_\mu ^0$ energy expected relative to the $%
\nu _e$ from three-body decay. Again, take 
\begin{equation}
R_D^{\prime }(M,U^2)=\frac{P_D^{\prime L_\mu ^0}(M,U^2)}{P_D^{\prime \nu
_{eD}}},
\end{equation}
with 
\begin{equation}
P_D^{\prime L_\mu ^0}(M,U^2)=\frac{U^2M^6\Delta }{\left\langle E_{L_\mu
^0}^{\prime }\right\rangle M_\mu ^5c\tau _\mu }\varepsilon _D^{L_\mu ^0}\eta
_D\left( \frac{M_\mu }M\right)
\end{equation}
and 
\begin{equation}
P_D^{\prime \nu _K}(M,U^2)=\left\langle \rho t\right\rangle \sigma _\nu
^{\prime }\left\langle E_{\nu _eD}\right\rangle N_A\varepsilon _D^{\nu _{eD}}
\end{equation}

The final  expression for $R_C(M,U^2)$ is 
\begin{equation}
R_C(M,U^2)=C^{\prime }U^4M^8\left( 1-\frac{M^2}{M_D^2}\right) ^2,
\end{equation}
with 
\[
C^{\prime }=a^{\prime }\frac{\eta _P(\frac M{M_D},\frac{M_\mu }M)\eta
_D\left( \frac{M_\mu }M\right) \varepsilon _D^{L_\mu ^0}\Delta }{%
\left\langle E_{L_\mu ^0}^{\prime }\right\rangle \left\langle E_{\nu
_{eD}}^{\prime }\right\rangle \varepsilon _D^{\nu _{eD}}\left\langle \rho
t\right\rangle },
\]
and 
\begin{eqnarray*}
a^{\prime } &=&\left[ M_\mu ^7\left( 1-\frac{M_\mu ^2}{M_D^2}\right) ^2c\tau
_\mu N_A\sigma _\nu ^{\prime }\frac{B(D^{+}\rightarrow \mu ^{+}X)}{%
B(D^{+}\rightarrow \mu ^{+}\nu _\mu )}\right] ^{-1} \\
&=&4.6\times 10^{13}\mbox{cm}^{-3}\mbox{g}^{-1}\mbox{GeV}^{-6}
\end{eqnarray*}
For fixed $M$, it follows that 
\begin{equation}
U_D^2\leq M^{-4}\sqrt{\frac 1{C^{\prime }}\left( 1-\frac{M^2}{M_D^2}\right)
^{-1}\frac{2.3}{f_{\nu _eD}f_{\nu _e}n_{\nu _\mu }N_{\mbox{POT }}}}.
\label{D-sense}
\end{equation}

Sensitivity extends to much lower mixing angle than the kaon case,
essentially because the $M^5$ behavior of the decay rate overwhelms the $%
\left( 1-\frac{M^2}{M_D^2}\right) ^2$ phase space factor, Charmed particles
are also typically produced with lower energy in a hadron collision than
kaons. This implies that $L_\mu ^0$ from charmed hadron decays will have
smaller time dilation factors, $E_{L_\mu ^0}^{\prime }/M$, and so the
probability that an $L_\mu ^0$ will decay in front of the detector is larger
if the $L_\mu ^0$ originates from a $D^{+}$ decay than it is for $L_\mu ^0$
from $K^{+}$ decay.

\subsection{$\mu $-type NHL from D$_S^{+}$ Mesons}

The limit on $U^2$ from this source follows directly from the result for $%
D^{+}$ mesons since one can immediately relate the number of NHL from $%
D_S^{+}$ decay to the number from $D^{+}$ decay via 
\begin{equation}
\frac{B(D_S^{+}\rightarrow \mu ^{+}L_\mu ^0)}{B(D^{+}\rightarrow \mu
^{+}L_\mu ^0)}=\frac{V_{cs}^2M_{D_S}(1-\frac{M^2}{M_{D_S}^2})^2f_{D_S}^2}{%
V_{cd}^2M_D(1-\frac{M^2}{M_D^2})^2f_D^2},
\end{equation}
and the ratio of produced $c\bar{s}$ mesons to $c\bar{d}$ mesons\cite{Charm
Production}, 
\begin{equation}
\frac{N(D_S^{+})}{N(D^{+})}=r_{D_SD}\simeq 0.5.
\end{equation}
It follows  that a limit on NHL produced from $D_S^{+}$ is directly related
to that from $D^{+}$ by 
\begin{equation}
U_{D_S}^2=U_D^2\sqrt{\frac{V_{cd}^2}{V_{cs}^2}\frac{M_D(1-\frac{M^2}{M_D^2}%
)^2}{M_{D_S}(1-\frac{M^2}{M_{D_S}^2})^2}\frac{f_D^2}{f_{D_S}^2}\frac 1{%
r_{D_SD}}}.  \label{Ds-sense}
\end{equation}
$D_S^{+}$ contribute more sensitivity to the rate than $D^{+}$ owing to the
Cabibbo favored annihilation diagram in the former's decay. This estimate's
main uncertainties are the ratio of $D_S^{+}$ to $D^{+}$ production, $%
r_{D_SD}$, which is taken to be $1/2$, and the ratio of pseudoscalar decay
constants, $f_D/f_{D_S}$, taken to be $1.0$.

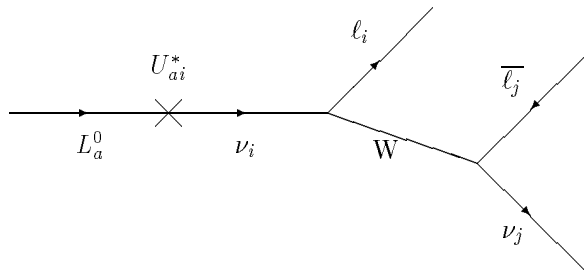
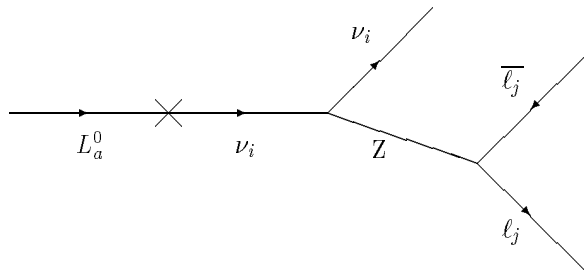
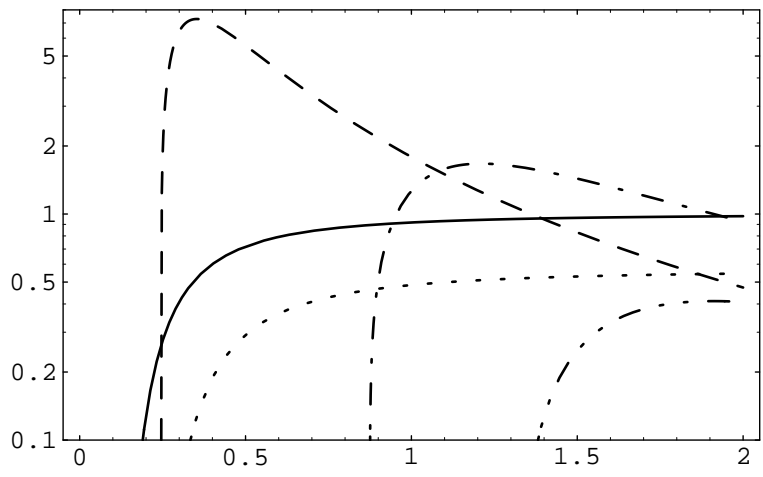
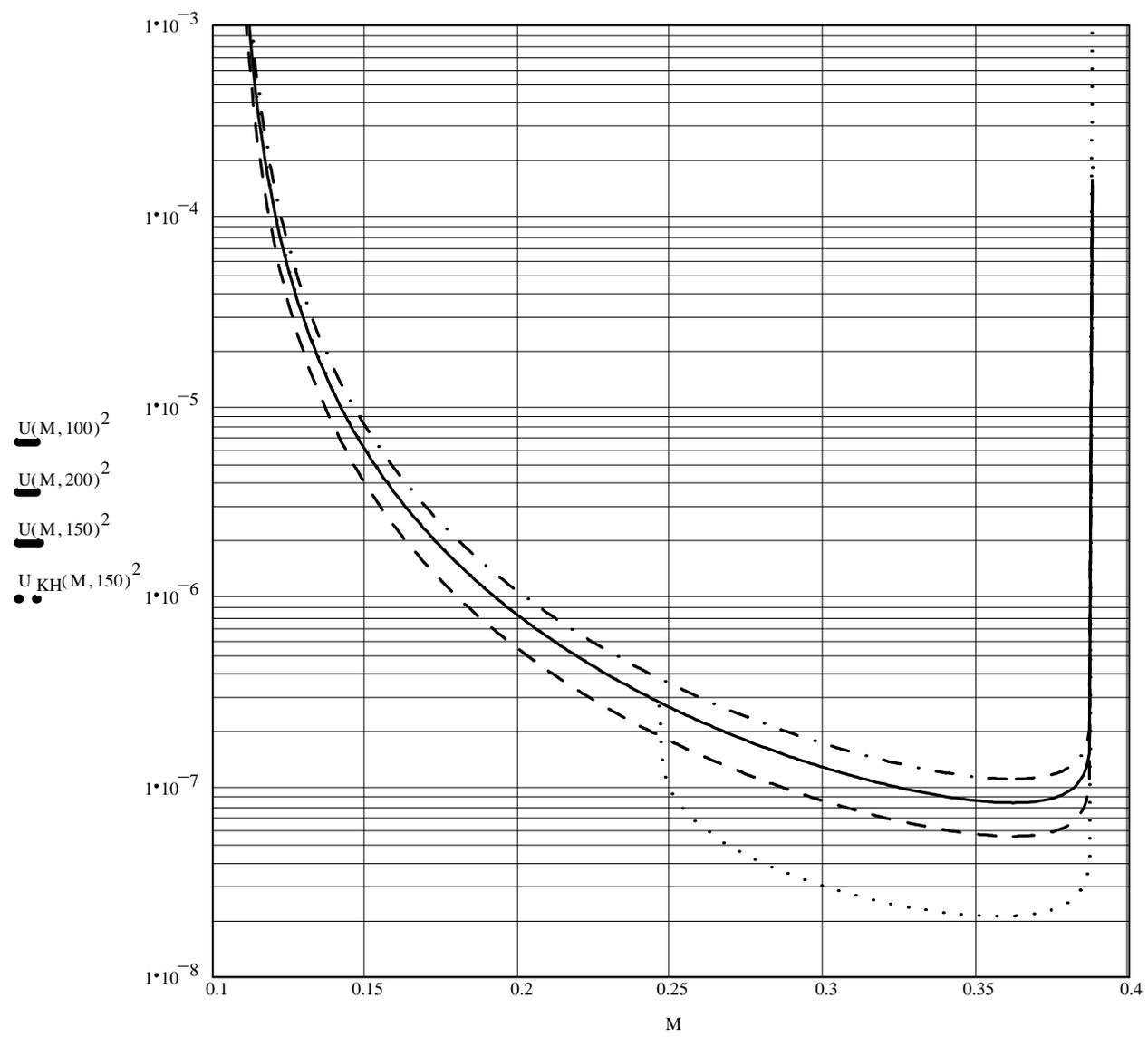
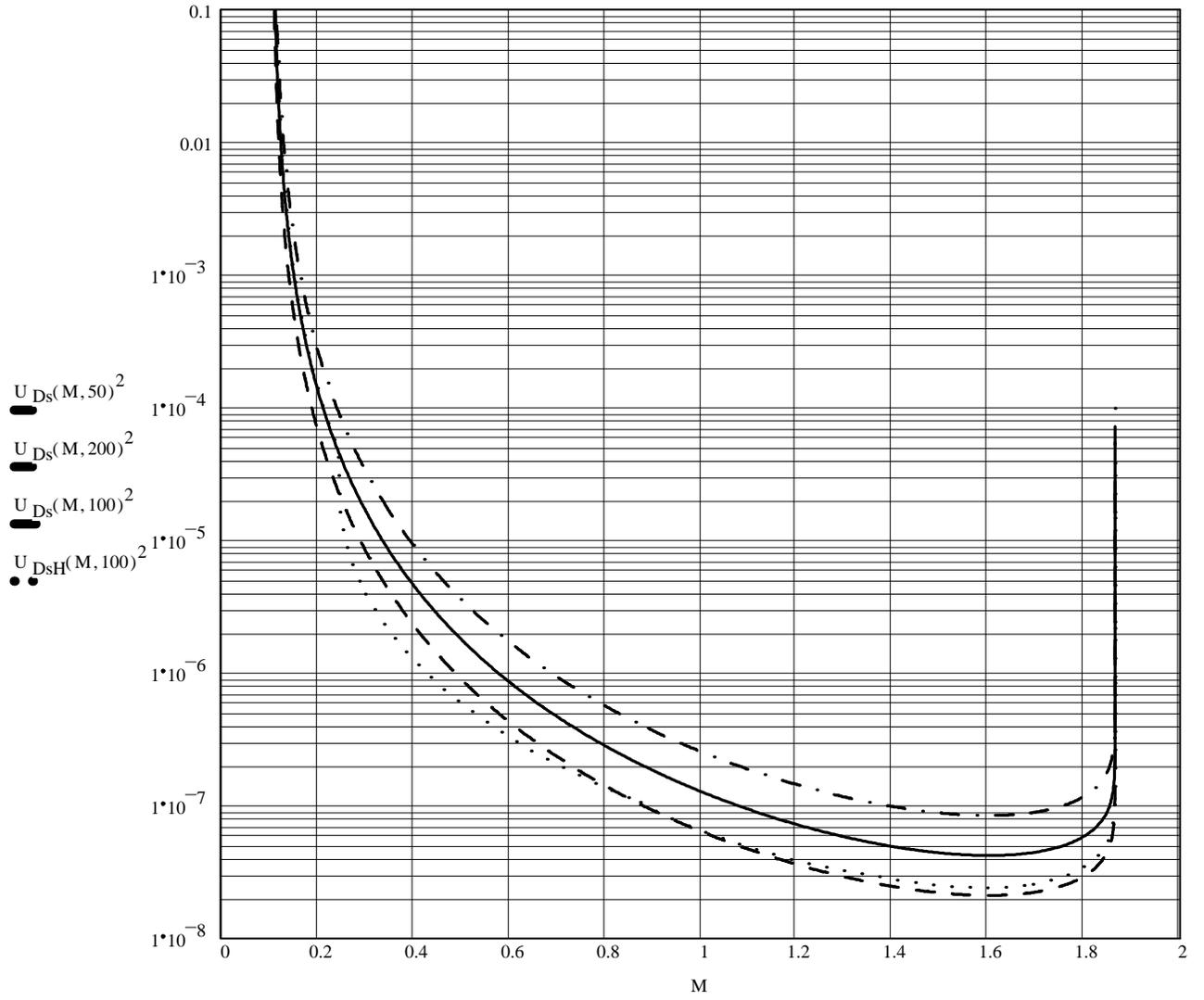
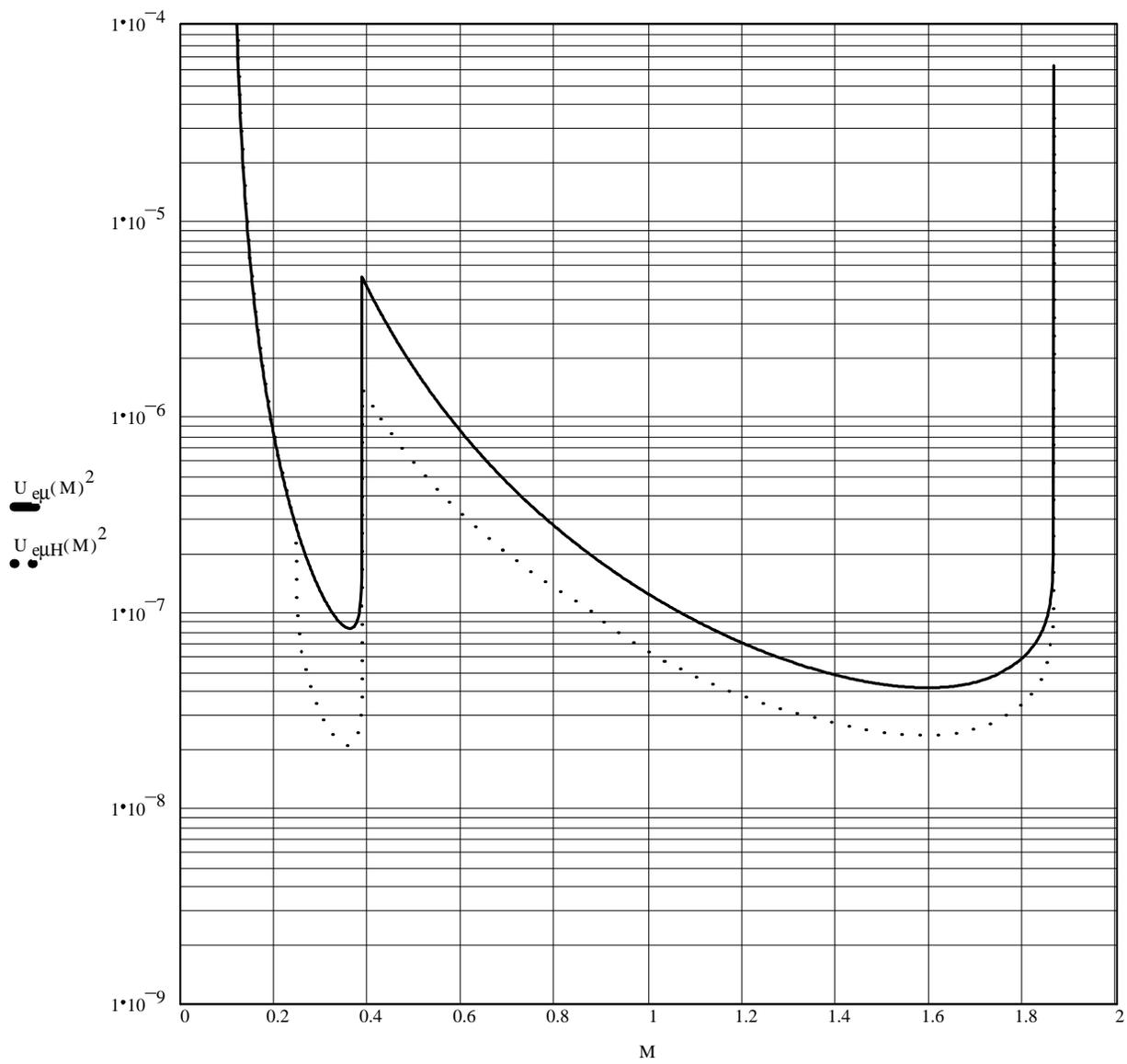
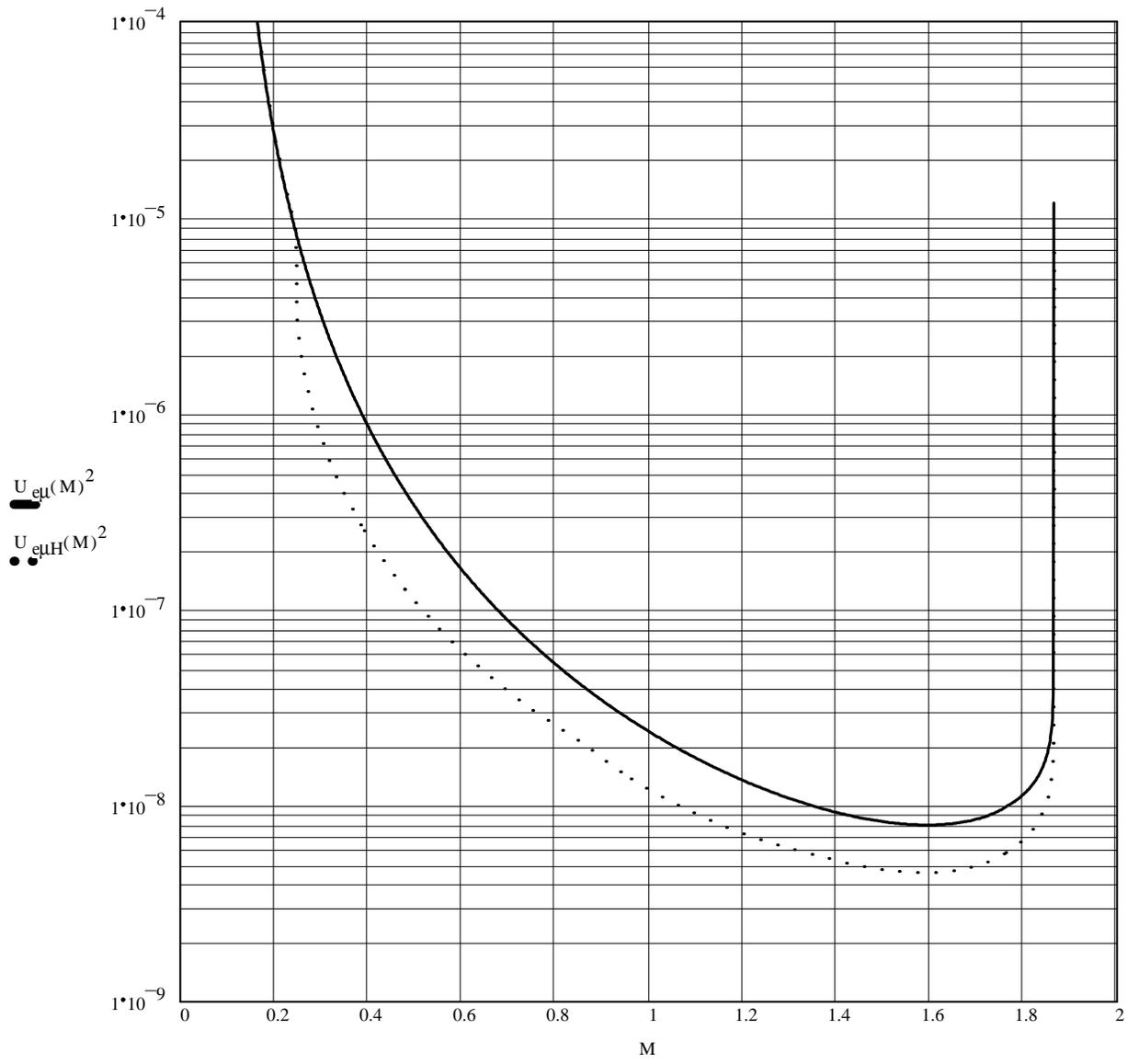
\end{document}